\documentclass{llncs}

\usepackage[hyphens]{url}
\usepackage[hidelinks]{hyperref}
\usepackage[utf8]{inputenc}
\usepackage{subcaption}
\usepackage{mathabx}
\usepackage{microtype}
\usepackage{multirow}

\usepackage{listings}
\usepackage{tikz}

\begin{document}

\title{Deep Linking Desktop Resources}

\author{
	Markus Schröder \and
	Christian Jilek \and
	Andreas Dengel
}
\institute{
	Smart Data \& Knowledge Services Dept., DFKI GmbH, Kaiserslautern, Germany\\ \and
	Computer Science Dept., TU Kaiserslautern, Germany\\
	\email{\{markus.schroeder, christian.jilek, andreas.dengel\}@dfki.de}
}

\maketitle

\begin{abstract}

Deep Linking is the process of referring to a specific piece of web content.
Although users can browse their files in desktop environments, they are unable to directly traverse deeper into their content using deep links.
In order to solve this issue, we demonstrate ``DeepLinker'', a tool which generates and interprets deep links to desktop resources, thus enabling the reference to a certain location within a file using a simple hyperlink. By default, the service responds with an HTML representation of the resource along with further links to follow.
Additionally, we allow the use of RDF to interlink our deep links with other resources.

\end{abstract}

\section{Introduction}

Internet resources are commonly identified using URIs\footnote{\url{https://tools.ietf.org/html/rfc3986}} which contain the information to look them up.
In this context, ``deep linking''\footnote{\url{https://www.w3.org/2001/tag/doc/deeplinking.html} (Section 2)} is the process of using a hyperlink to link to a specific piece of web content.
For example, \url{https://en.wikipedia.org/wiki/Deep_linking#Example} refers to the Example section of Wikipedia's Deep Linking page.
Similarly, desktop resources, like files (in various formats\footnote{In addition to files, other resources may reside on the desktop as services like, for example, mails (IMAP), databases (SQL) or on other external endpoints.}) and folders, are typically addressed with their absolute paths or file URI schemata\footnote{\url{https://tools.ietf.org/html/rfc8089}}.
Although users can browse a file in this way, they are unable to directly descend deeper into its content.
As a result, users cannot refer to a certain location deep inside their desktop resources,
for example, referring to a shape in a slide of a presentation.
This is due to a missing definition and standardisation, and as a consequence, absent implementations of common desktop applications such as editors or readers.

Thus, in this paper we demonstrate ``DeepLinker'' which allows browsing desktop resources to an arbitrary depth using deep links.
The core idea is to generate, interpret and maintain the URIs referring to a certain location within desktop resources as well as showing an HTML representation of the browsed fragment in order to provide a visualization.
For better demonstration an online prototype is available\footnote{\url{www.dfki.uni-kl.de/~mschroeder/demo/deeplinker}}.

\section{Related Work}
\label{sec:relwork}

To satisfy the need for identifying subordinate resources in the web, the URI standard defines fragment identifiers\footnote{\url{https://tools.ietf.org/html/rfc3986\#section-3.5}}.
However, usual desktop applications are not aware of the fragment identifier concept.
One would have to reimplement or extend all of them (e.g. with plug-ins) to enable a similar behaviour.
That is why we decided to write our own application.
To simulate a Desktop application rendering resources we convert and present them using HTML.

Other approaches like the LEMO annotation framework \cite{haslhofer2009lemo} use fragment identification for MPEG resources \cite{mpeg} as well as other fragment identifiers.
A 2007 survey \cite{Jochum2007} showed that fragmentation links for complex documents (e.g. spreadsheets, charts, presentations and word processing documents) do not exist.
In fact, current media fragments\footnote{\url{https://www.w3.org/TR/media-frags/}} solely focus on image, audio and video.

\section{DeepLinker}
\label{sec:approach}

Our tool generates and interprets deep links to desktop resources, thus making it possible to refer to a certain location within a file using a simple hyperlink. An HTML page presents the referred part to the users.

\begin{figure}[p]
	\centering
	\begin{subfigure}[t]{0.45\linewidth}
		\caption{\scriptsize{\href{http://173.212.240.179:7276/filesystem/pictures/a.png/content/to@image/rect@600,109,188,36}{\texttt{/filesystem/a.png/content/to@image\\/rect@600,109,188,36\\}}}}
		\fbox{\includegraphics[width=\textwidth]{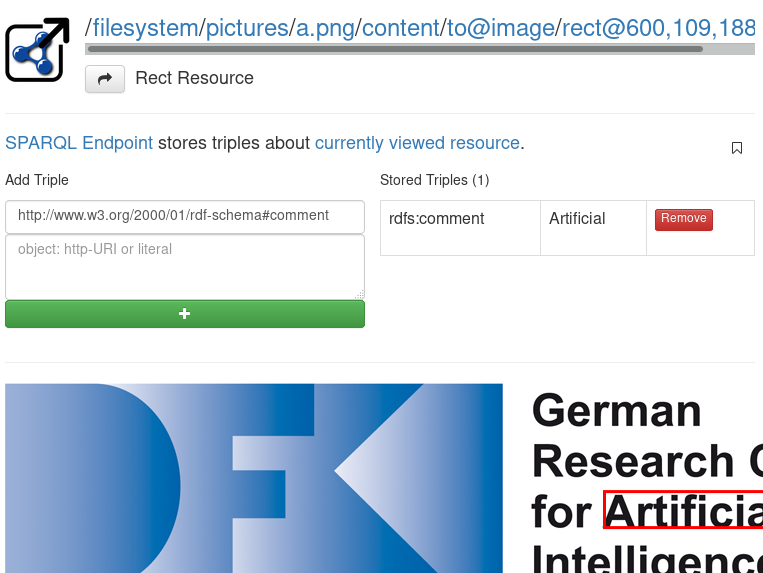}}
		\label{fig:01}
	\end{subfigure}
	\quad
	\begin{subfigure}[t]{0.45\linewidth}
		\caption{\scriptsize{\href{http://173.212.240.179:7276/filesystem/presentations/b.pptx/content/to@powerpoint/index@3/cssSelector@svg\%2B\%253E\%2Bg\%2B\%253E\%2Bg\%253Anth-child\%252843\%2529}{\texttt{/filesystem/b.pptx/content/to@powerpoint\\/index@3/cssSelector@svg\%2B\%253E\%2Bg\%2B\%253E\\\%2Bg\%253Anth-child\%252843\%2529}}}}
		\fbox{\includegraphics[width=\textwidth]{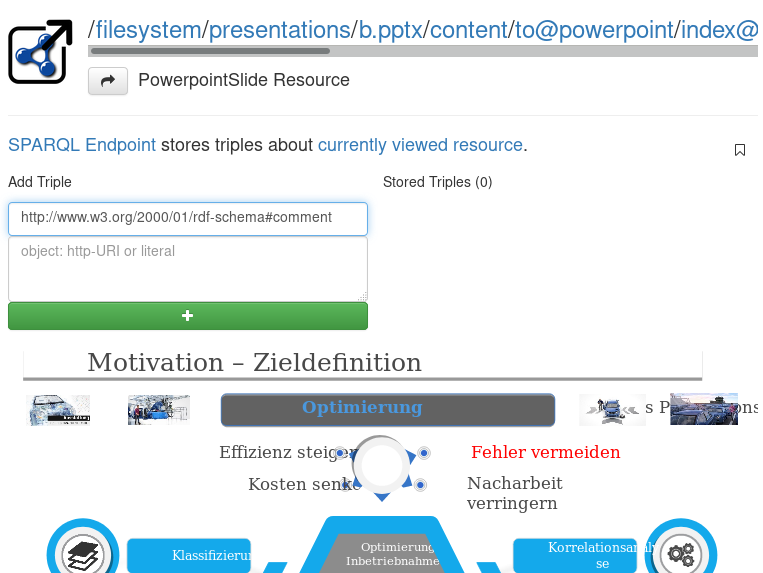}}
		\label{fig:02}
	\end{subfigure}
	\quad
	\begin{subfigure}[t]{0.45\linewidth}
		\caption{\scriptsize{\href{http://173.212.240.179:7276/filesystem/c.txt/content/to@string/line@2}{\texttt{/filesystem/c.txt/content/to@string\\/line@2\\\quad}}}}
		\fbox{\includegraphics[width=\textwidth]{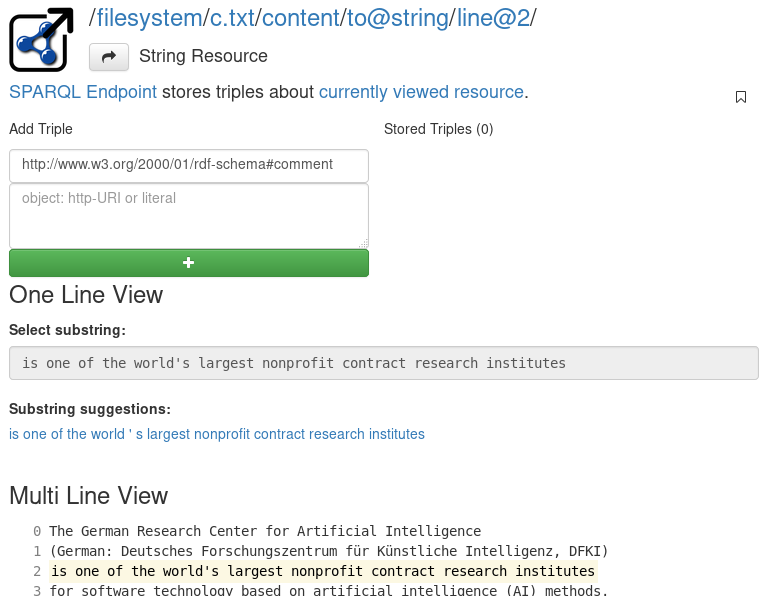}}
		\label{fig:03}
	\end{subfigure}
	\quad
	\begin{subfigure}[t]{0.45\linewidth}
		\caption{\scriptsize{\href{http://173.212.240.179:7276/remote/download@http\%253A\%252F\%252Fw3c.org,*\%252F*/content/to@html/cssSelector@\%2523w3c_nav\%2520\%253E\%2520form\%253Anth-child(2)\%2520\%253E\%2520ul.main_nav\%2520\%253E\%2520li\%253Anth-child(2)\%2520\%253E\%2520a}{\texttt{/remote/download@http\%253A\%252F\%252F\\w3c.org,*\%252F*/content/to@html/cssSelector@\\\%2523w3c\_nav\%2520\%253E\%2520form\%253\dots}}}}
		\fbox{\includegraphics[width=\textwidth]{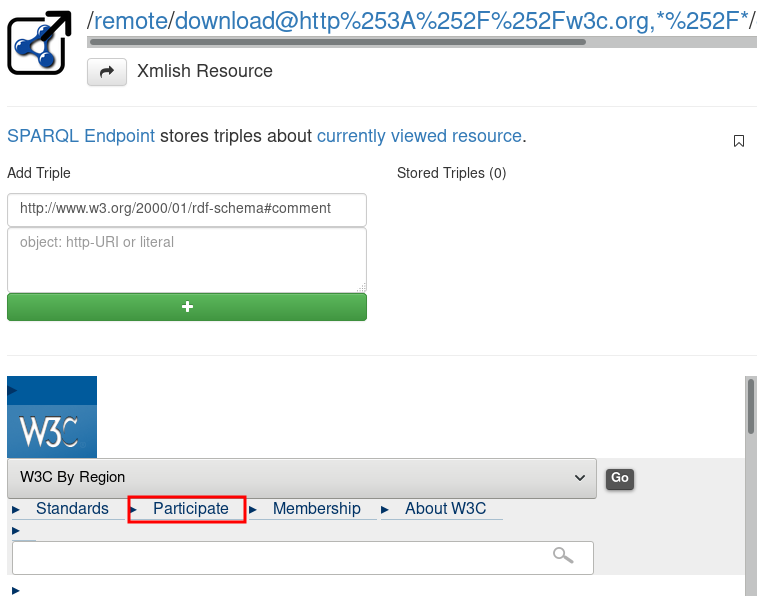}}
		\label{fig:04}
	\end{subfigure}
	\caption{
		(a) Focusses the word ``Artificial'' in DFKI's logo which is annotated with an \texttt{rdfs:comment} ``Artificial'',
		(b) highlights the shape ``Fehler vermeiden'' in the 4th slide of the Powerpoint presentation \texttt{b.pptx}, 
		(c) selects the 3rd line of a text file, 
		(d) refers to the ``Paticipate'' element in the downloaded \href{http://w3c.org}{\texttt{w3c.org}} page.
	}
	\label{fig:imgs}
\end{figure}

For a first impression, \autoref{fig:imgs} exemplifies four deep links together with the web sites they refer to.
In general, the pages show the accessed link and a simple form to add and list RDF triples below.
The images depict the following cases each having a different highlighting:
(a) a part of an image,
(b) a shape in a presentation slide (text in red),
(c) a line in a text file, and
(d) an element in a web page.

\begin{figure}[p]
	\centering
	\includegraphics[width=\textwidth]{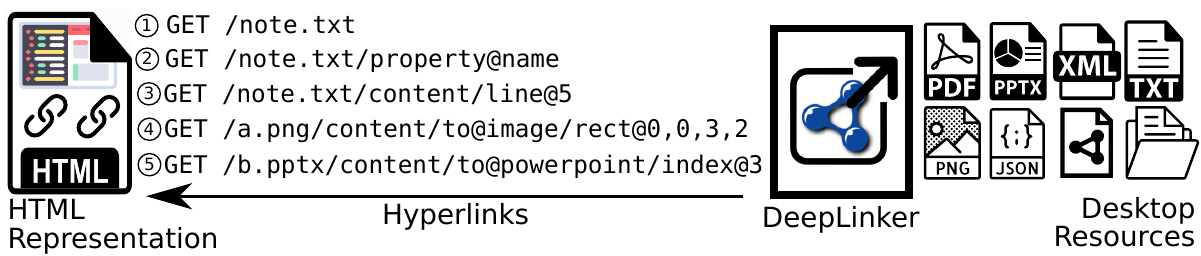}
	\caption{
		DeepLinker browses usual desktop resources by using various path segments. Usually, our app returns an HTML representation of the fragment with further links to follow. Hyperlinks:
		(1) returns file information of \texttt{note.txt} (not its content),
		(2) presents the name of the file,
		(3) highlights its 5th line.
		(4) draws a rectangle on image \texttt{a.png}.
		(5) shows the 3rd slide of \texttt{b.pptx}.
	}
	\label{fig:architecture}
\end{figure}

\autoref{fig:architecture} shows the environment in which DeepLinker is used together with additional example links.
Usually, our tool runs locally on the users' PCs.
A DeepLinker resource is requested by using its hyperlink (deep link) in a web browser.
Our tool responds with an HTML representation of the resource along with links pointing deeper into the file.
As a result, instead of just stopping at the resource's content (surface link), users can traverse further into the resources enabling a more fine-grained selection of a desired fragment.


Our approach allows for a hierarchical drill-down using the URI path segments.
The deep links are processed by the server at runtime in the following way:
Each path segment is implemented as a parametrized method returning a DeepLinker resource.
Thus, they can be chained in order to traverse arbitrarily deep.
The segments are serialised in the URI in the following way: \texttt{/<method>@<param1>,...,<paramN>/}.

For demonstration purpose we implemented nine path segment methods along with 14 resource types.

\subsubsection{Path Segment Methods.}

The \texttt{child}\footnote{If no method is given, \texttt{child} is assumed to be the default.} and \texttt{index} methods are convenient ways to select a sub-resource by name or sequence number.
\texttt{line} and \texttt{substring} select corresponding parts of a string.
\texttt{rect} is meant to highlight a part of an image, while \texttt{cssSelector} is used to refer to an element in an XML-like structure.
\texttt{download} retrieves an external file using its URL.
In order to acquire a value for a given key the \texttt{property} method can be used.

In contrast to the previous read-only methods, the \texttt{to} method allows for transforming resources from one format to another (if this functionality is available for the respective DeepLinker resource type).

\subsubsection{Resource Types.}

Currently 14 resource types are supported.
\texttt{Collection} and \texttt{Map} store resources the same way known from programming languages.
\texttt{File} holds a local file's meta data but not its content.
The latter is represented by \texttt{String} (plain text), \texttt{JSON}, \texttt{Image}, \texttt{PDF}, \texttt{Powerpoint(Slide)} and \texttt{RDF} depending on its associated format.
The raw content can also be represented using the \texttt{Binary} resource type.
A special \texttt{Rect} type models the actual highlighted area.
\texttt{Xmlish} is a more general resource type covering XML and XML-based structures like HTML or SVG.
The \texttt{Remote} resource serves as an entry point to upload and download files.

\subsubsection{Annotations via RDF.}

Like stated before in more detail, DeepLinker generates and interprets deep links to desktop resources.
Having these links now enables us to make statements about them using RDF.
This approach is comparable with the Annotea project \cite{koivunen2005annotea} which stores meta data about websites.
Similarly, Deeplinker additionally allows to store and retrieve RDF using an external SPARQL endpoint.
For demonstration purposes, DeepLinker is equipped with a Fuseki endpoint running localhost at \texttt{/fuseki/annotation}.
In the resource's HTML representation, our prototype provides a form to add RDF statements and inspect stored ones. 
With this capability it is possible to annotate any desktop resource (and especially their fragments) with further meta data.
The resources' literals are searchable in order to find associated DeepLinker links.
In order to simulate bookmarking, users may conveniently add a triple of the form \texttt{<DeepLinkerLink> rdf:type} \url{https://www.w3.org/2002/01/bookmark\#Bookmark} with a single click.
Bookmarks are queried and listed using a separate resource page.

\subsubsection{Content Negotiation.}

By default, DeepLinker returns an HTML page rendering the resource and providing further links.
Users may thus browse their desktop resources in a familiar way.
However, DeepLinker supports content negotiation based on the provided accept header in the request (if the requested resource type implements it).
For example, given the accept header \texttt{application/json} returns the resource serialized as JSON.
In case of \texttt{text/turtle}, the resource is converted to RDF and serialized in Turtle format (currently only implemented for file resources).
Following the third Linked Data principle\footnote{\url{https://www.w3.org/DesignIssues/LinkedData.html}}, additional statements about the resource queried using the SPARQL endpoint are added, too.

\section{Conclusion and Outlook}
\label{sec:concl}

In this paper, we demonstrated DeepLinker, a tool that allows for browsing usual desktop resources arbitrarily deep, enabling users to refer to any desired fragment.
This is accomplished by generating and interpreting deep links.
Currently, our prototype implements nine path segment methods together with 14 resource types, especially making it possible to use RDF to annotate the now existing links with further meta data.
Our prototype can be tested online\footnote{\url{www.dfki.uni-kl.de/~mschroeder/demo/deeplinker}}.

In the future, we think of supporting data scientists in business and data understanding phases of data mining processes.
Using our tool they would then be able to browse and annotate their database content and CSV files in order to store newly acquired knowledge.
In this regard, we also think of collaborative scenarios in which users create and share deep links among each other.

\bibliographystyle{llncs}
\bibliography{paper}

\end{document}